\documentstyle[11pt,equation,
epsf]{article}
\input{psfig.sty}
\begin{document}
\newcommand{\tr}{\rm Tr}
\newcommand{\sx}{\sigma}
\newcommand{\sxa}{\sigma_1}
\newcommand{\sxb}{\sigma_2}
\newcommand{\pha}{\phi_1}
\newcommand{\phb}{\phi_2}
\newcommand{\phpa}{\phi^+_1}
\newcommand{\phpb}{\phi^+_2}
\newcommand{\phm}{\phi^-}
\newcommand{\php}{\phi^+}
\newcommand{\psif}{{\psi_f}}
\newcommand{\psa}{\psi_1}
\newcommand{\psb}{\psi_2}
\newcommand{\Phib}{\bar{\Phi}}
\newcommand{\mpl}{m_{Pl}}
\newcommand{\Mpl}{M_{Pl}}
\newcommand{\lx}{\lambda}
\newcommand{\Lx}{\Lambda}
\newcommand{\kx}{\kappa}
\newcommand{\ex}{\epsilon}
\newcommand{\be}{\begin{equation}}
\newcommand{\ee}{\end{equation}}
\newcommand{\een}{\end{subequations}}
\newcommand{\ben}{\begin{subequations}}
\newcommand{\beq}{\begin{eqalignno}}
\newcommand{\eeq}{\end{eqalignno}}
\def \lta {\mathrel{\vcenter
     {\hbox{$<$}\nointerlineskip\hbox{$\sim$}}}}
\def \gta {\mathrel{\vcenter
     {\hbox{$>$}\nointerlineskip\hbox{$\sim$}}}}
\pagestyle{empty}
\noindent
\begin{flushright}
SNS--PH/1998--5
\\
INFN-FE 03-98 
\\
hep--ph/98xxxxx
\end{flushright} 
\vspace{3cm}
\begin{center}
{ \Large \bf
Field Evolution Leading to Hybrid Inflation
} 
\\ \vspace{1cm}
{\large 
Z. Berezhiani$^{(a,b)}$, D. Comelli$^{(a)}$ and N. Tetradis$^{(c)}$ 
} 
\\
\vspace{1cm}
{\it
(a) INFN, Sezione di Ferrara, Via Paradiso 12, 44100 Ferrara, Italy 
\\
(b) Institute of Physics, Georgian Academy of Sciences, Tbilisi, Georgia
\\
(c) Scuola Normale Superiore, Piazza dei Cavalieri 7, 56100 Pisa, Italy 
} 
\\
\vspace{2cm}
\abstract{
In general the onset of hybrid inflation requires 
an extremely homogeneous field configuration at the end of
the Planck era.
The field $\phi$ orthogonal to the inflaton $\sx$ must be zero
with high accuracy 
over a range 
exceeding the initial horizon size $\sim \mpl^{-1}$
by about two orders of magnitude. 
We consider a supersymmetric model that 
permits the decay of the oscillating field $\phi$
into light particles. We study the field evolution 
and find that 
the requirement of extreme homogeneity can be relaxed.
However, the field $\phi$ must still be 
smaller than the inflaton 
by a factor of order 1
over a region far exceeding the initial
horizon size. 
\\
\vspace{1cm}
} 
\end{center}
\vspace{4cm}
\noindent
March 1998

\newpage

\pagestyle{plain}
\setcounter{page}{1}

\setcounter{equation}{0}

\section{Initial conditions for hybrid inflation} 

An attractive realization of the inflationary scenario is obtained
when hybrid inflation \cite{hybrid}
is embedded in the context of supersymmetry 
\cite{cop}--\cite{gia}.
Flat directions with non-zero potential energy density
appear without fine-tuning and
are not lifted by quantum corrections.
The COBE observation of the cosmic microwave background
anisotropy constrains the properties of the model along the 
inflationary trajectory \cite{star}--\cite{report}.
On general grounds, one expects the inflationary energy scale
$V^{1/4}$ 
(determined by the vacuum energy density during inflation) 
to be at least two or three orders of magnitude smaller
than the Planck scale 
\footnote{We refer to the ``reduced'' Planck scale 
$\mpl= \Mpl/\sqrt{8 \pi}$, $\Mpl = 1.22 \times 10^{19}$ GeV.}
\cite{lyth}:
\be
V^{1/4} / \ex^{1/4} \simeq 7 \times 10^{16}~{\rm GeV}.
\label{oneone} \ee
Here $\ex$ is a ``slow-roll'' parameter \cite{report} that 
must be much smaller than 1 during inflation.

The question of initial conditions for hybrid inflation
has been addressed recently \cite{first1}--\cite{nikos}.
In general, the onset of inflation requires a region of
space with a size of a few Hubble lengths
where the fields take almost constant values, so that the
gradient energy density is negligible compared to the potential energy 
density \cite{gold}. 
For hybrid inflation, the requirement is much stricter \cite{nikos}. 

Let us consider a supersymmetric model of hybrid inflation 
\cite{cop,shafi} based on the renormalizable superpotential 
\be
W =   - \mu^2 S  + \lx S \Phi^2.
\label{one1} \ee
It is invariant under a global $R$-symmetry with the
superfields $S$ and $\Phi$ carrying $R$-charges 1 and 0 respectively. 
The scalar components of these superfields 
can be written in the form 
\be
S =\frac{\sx}{\sqrt{2}},~~~~~~~~~~~~~
\Phi = \frac{\pha +i \phb}{\sqrt{2}}. 
\label{um1} \ee
Here we have used 
an $R$-transformation in order to make $S$ real.  
The potential 
is given by the expression
\be
V = \mu^4 -\lx \mu^2 \left( \pha^2 - \phb^2 \right) 
+\frac{\lx^2}{4} \left( \pha^2 + \phb^2 \right)^2 
+ \lx^2 \sx^2 \left( \pha^2 + \phb^2 \right).
\label{three1} \ee 

The supersymmetric vacua are located at
$\sx=0$, $\pha^2= 4\mu^2/\lx$, $\phb=0$. 
The potential has a flat direction along the $\sx$ axis, 
for $\sx > \sx_{ins} = \mu/\sqrt{\lambda}$, 
with the non-zero vacuum energy 
$V(\sx,\pha=\phb=0)=\mu^4$ that can support inflation. 
The flatness of the potential along the $\sx$ axis is lifted by 
radiative corrections, resulting from the breaking of supersymmetry 
by the non-zero value of $V(\sx,\pha=\phb=0)$. 
The precise value of these corrections 
is not important for our discussion. They induce the slow roll of
the inflaton field once inflation has set in, but have a very small
effect on the question of the onset of inflation. For this reason 
we neglect them in the following. 

We assume that a Robertson-Walker metric is a good approximation 
for the regions of space with uniform fields that we are considering. 
The evolution of the fields is given by the standard equations
\beq
\ddot{\sx} + 3 H \dot{\sx} = &~ 
- \frac{\partial V}{\partial\sx},
\label{twoeight} \\
\ddot{\phi}_{1,2} + 3 H \dot{\phi}_{1,2} = &~ 
- \frac{\partial V}{\partial\phi_{1,2}}, 
\label{twoeighta} \\
H^2 = \left( \frac{\dot{R}}{R} \right)^2 = &~\frac{1}{3 \mpl^2} \left(
\sum_i \frac{1}{2} \dot{\varphi}_i^2 
+ V \right). 
\label{twonine} \eeq

Near the onset of inflation, 
the fields $\phi_{1,2}$ oscillate around zero with amplitudes 
$\left[ \phi_{1,2} \right]$. The equation of motion for $\sx$
reads 
\be
\ddot{\sx} + 3 H \dot{\sx} =  
- \lx^2
\left( \left[ \pha \right]^2 + \left[ \phb \right]^2
\right) \sx,
\label{twothirteen} \ee
where we replaced $\pha^2$ and $\phb^2$
by their
average values $\langle \phi_{1,2}^2 \rangle
\sim \frac12\left[ \phi_{1,2} \right]^2$ during an oscillation.

Two time scales
characterize the solutions of this equation.
The first one is related to the ``friction'' term and is given by  
$t_H^{-1} = 3H/2$.
Near the onset of inflation, the Hubble parameter is dominated
by the non-zero vacuum energy $V=\mu^4$, where the scale $\mu$ is
constrained by the COBE observations to be two or three orders of
magnitude below $\mpl$. This gives
\be
t_H =  \frac{2}{\sqrt{3}}  \frac{\mpl}{\mu^2}.
\label{twofourteen} \ee
The other time scale is obtained if we neglect the ``friction'' term and 
consider the oscillations of the $\sx$ field.
One-fourth of the period is the typical time for the system to roll
to the origin and away from an inflationary solution. It is given
by 
\be 
t_{osc} = \frac{\pi}{2 \lx} 
\frac{1}{\sqrt{\left[ \pha \right]^2 + \left[ \phb \right]^2}}.
\label{twofifteen} \ee

Inflation sets in if $t_{osc} \gta t_H$ and the $\sx$ field
settles at a constant value on the $\sx$ axis.  
This requirement can be written as
\be
\frac{\sqrt{\left[ \pha \right]^2 + \left[ \phb \right]^2}}{\mu} 
\lta \frac{\sqrt{3}~\pi}{4~\lx} 
\frac{\mu}{\mpl} .
\label{twosixteen} \ee
For a scale $\mu$ consistent with the 
COBE observations ($\mu/\mpl \sim 10^{-3}$--$10^{-2}$) and couplings of
order 1, 
the value of the fields $\phi_{1,2}$ orthogonal to the
inflaton must be zero with an accuracy of two or three orders of 
magnitude in units of $\mu$.
This condition of extreme homogeneity must be satisfied 
throughout a region of space with a size of the order of the  
Hubble length $\sim \mpl/\mu^2$. Otherwise, the fields in
different parts of this 
space region will evolve towards very different values.
In one part they may settle in the valley along the inflaton axis,
while in another they may end up at the minima of the potential. 
As a result, large inhomogeneities will appear at scales smaller than
$\sim H^{-1}$ that will prevent the onset of inflation \cite{nikos}. 

On the other hand, 
the earliest time at which one could start considering 
regions of space with classical fields 
is when the Planck era ends and 
classical general relativity becomes applicable. 
The initial energy density is near $\mpl^4$. 
For a theory with couplings
not much smaller that 1, 
the initial field values within each region, 
as well as the initial field fluctuations,
are expected to be of order $\mpl$.
It is very difficult for such an initial field
configuration to evolve into the extremely homogeneous configuration 
that is required at the onset of hybrid inflation. 

In fig. 1 we present the evolution of 
the fields $\sx$, $\pha$ 
for a theory described by
the potential of eq. (\ref{three1}) 
with $\lx=0.3$, $\mu/\mpl=10^{-3}$.
The initial values of the fields have been chosen
$\sx/\mpl=3$,
$\pha/\mpl=0.6$,
$\phb/\mpl=0.5$.
They are near $\mpl$, with 
$\pha$, $\phb$ smaller than $\sx$, so that the evolution starts
near the flat direction.
The initial
time derivatives of the fields have been set to zero and 
the evolution equations (\ref{twoeight})--(\ref{twonine})
have been integrated numerically. 
We observe an initial stage during which the $\sx$ field is reduced, while
$\pha$ and $\phb$ oscillate rapidly around zero with quickly decaying
amplitudes. This quick decay is due to
the large initial value of $H$, induced by the large average value
of $V$ in eq. (\ref{twonine}).
When the fields become much smaller than their initial values, $H$ 
drops significantly and the ``friction'' term in eqs. (\ref{twoeight}),
(\ref{twoeighta}) becomes less important. The amplitudes of the 
$\phi_{1,2}$ fields are much larger than the bound of eq.(\ref{twosixteen}). 
As a result, $\sx$ rolls below zero and into negative values. 
Eventually, the system of fields
enters a stage of oscillations around zero with slowly
decaying amplitudes. The frequency of oscillations of 
$\phi_{1,2}$ is larger than that of $\sx$ because of the
difference of their effective masses induced by the last term 
in the right-hand side of eq. (\ref{three1}).

It is clear from the above discussion that the onset of
hybrid inflation requires the fine-tuning of the
initial field configuration in such a way that the amplitudes of
$\phi_{1,2}$ are smaller than the bound of eq. (\ref{twosixteen}) when
the energy density has dropped from $\sim \mpl^4$
to $\sim \mu^4$.
In general, this leads to a requirement of extreme homogeneity 
for the initial field configuration at the end of the Planck
era (with an accuracy of up to ten orders of magnitude in 
units of $\mpl$) -- an "unnatural'' assumption  \cite{nikos}.

What should be the initial scale of the  
homogeneity  at the end of the Planck era? 
The size $L$ of a homogeneous region evolves 
proportional to the scale factor and, in terms of the Hubble length,  
is given by 
\be
\frac{LH}{L_0H_0} \sim \left( \frac{\rho}{\rho_0} 
\right)^{\frac{1+3w}{6(1+w)}}.
\label{fin} \ee
The parameter $w$ determines the relation between energy density and
the mean value of the pressure ($p=w\rho)$. 
For a system of massive oscillating fields, or a 
matter-dominated Universe, $w = 0$, and for a 
radiation-dominated Universe $w = 1/3$. 
At the onset of inflation, when
$\rho/\mpl^4 \sim \mu^4/\mpl^4 \sim 10^{-8}$--$10^{-12}$, 
we must have $LH \sim 1$. For this we must 
require $L_0 H_0 \gta 10$--$100$
at the end of the Planck era, when $\rho_0/\mpl^4 \sim 1$.   
This means that the initial extreme homogeneity must extend 
far beyond the typical size of regions that can be considered
causally connected.

A feature of the field evolution that was not taken into account
in ref. \cite{nikos} concerns the possible decay of the 
oscillating fields $\phi_{1,2}$ into light particles
\cite{dolgov}. As these fields
are coupled to the inflaton, they can produce $\sx$ particles. 
However, the production rate through the coupling 
$\lx^2 \sx^2 \left( \pha^2 + \phb^2 \right)$
is proportional to the square of the amplitude of the 
$\phi_{1,2}$ oscillations and becomes negligible very quickly. As 
a result, this particle production 
does not alter the picture we sketched above.

The possibility exists, however, that the 
oscillating fields $\phi_{1,2}$ produce other light particles
with a large decay rate. 
In this letter we consider a supersymmetric
model for which such a scenario is realized, and  
show that in this situation the requirement of the 
extreme homogeneity can be relaxed. Namely, provided 
that the preinflationary reheeting is effective,  
inflation can proceed from the space region in which 
after the Planck era the initial field values 
are $\phi_{1,2}\lta \sx \sim \mpl$.  
However, still the size of this region should be 
larger than the initial horizon size $\sim \mpl^{-1}$ 
by about two orders of magnitude.

\section{Particle production before the onset of inflation} 

We consider the renormalizable superpotential 
\be
W =   - \mu^2 S  + \lx S \Phi^2 + \kx \Phi \Psi^2 + m \Psi^2.
\label{one} \ee
It is invariant under a global $R$-symmetry with the
superfields $S$, $\Phi$, $\Psi$ carrying 
$R$-charges 1, 0, 1/2 respectively. We have not included a term
$M S \Phi$, even though it is allowed by the $R$-symmetry,
because it can be eliminated through a redefinition of the superfields. 
Moreover, this term is absent if the 
superfields $\Phi$ and $\Psi$ belong in non-trivial real 
representations of some symmetry group under which $S$ is a singlet.
We assume that the two mass scales $\mu$, $m$ are comparable and
the couplings $\lx$, $\kx$ not much smaller than 1.

The scalar components of the superfields can be 
written in the form 
\be
S =~\frac{\sx}{\sqrt{2}},
~~~~~~~~~~~~
\Phi = ~\frac{\pha +i \phb}{\sqrt{2}},
~~~~~~~~~~~
\Psi = ~\frac{\psa +i \psb}{\sqrt{2}}.
\label{two} \ee
Here we have used an $R$-transformation in order to make $S$ real.
The potential is given by the expression
\beq
V=&~
\mu^4
-\lx \mu^2 \left( \pha^2 - \phb^2 \right) 
+\frac{\lx^2}{4} \left( \pha^2 + \phb^2 \right)^2 
\nonumber \\
&
+ \left( \psa^2 + \psb^2 \right)
\left[
2m^2 + \kx^2 \left( \pha^2 + \phb^2 \right)
+2 \sqrt{2}\kx m \pha
\right]
\nonumber \\
&
+ \lx^2 \sx^2 \left( \pha^2 + \phb^2 \right)
+\frac{\kx^2}{4} \left( \psa^2 + \psb^2 \right)^2 
+\kx \lx \sx \left[ 
\pha \left( \psa^2 - \psb^2 \right)
+2 \phb  \psa \psb
\right].
\label{three} \eeq 

The supersymmetric vacua are located at
$\sx=0$, $\pha^2= 4\mu^2/\lx$, $\phb=0$, $\psa=0$, $\psb=0$. 
The potential has a flat direction (independent of $\sx$)
for $\pha=\phb=\psa=\psb=0$.
Along the flat direction the masses of the 
$\phi_{1,2}$ fields are given by 
\beq
M^2_{\pha}=~&2\lx^2 \sx^2- 2\lx\mu^2, 
\nonumber \\
M^2_{\phb}=~&2\lx^2 \sx^2 +2\lx\mu^2.
\label{four} \eeq
An instability appears for 
\be
\sx^2 
< \sx^2_{ins} =  \frac{\mu^2}{\lx},
\label{five} \ee
which can trigger the growth of the $\pha$ field.

The coupling $\Phi \Psi^2$ permits the decay of the $\phi_{1,2}$ fields
into $\Psi$ particles. 
If these fields perform coherent oscillations, 
they can decay into scalar or fermionic $\Psi$ particles. 
In the most favourable situation, $\sx$ is slowly varying
in the region $\sx \gg \sx_{ins}$ and the frequency of oscillations
is $\sqrt{2} \lx \sx \gg 2 m$. 
The decay rates of each one of the $\phi_{1,2}$ fields 
into scalar or fermionic $\Psi$ particles are 
equal 
\footnote{In our model the fermionic $\Psi$ particles are
Majorana spinors.} and
\beq
\Gamma=&~\Gamma_s+\Gamma_f , 
\nonumber \\
\Gamma_s=&~\Gamma_f = \frac{1}{4\sqrt{2}\pi}\kx^2\lx\sx.
\label{six} \eeq

The evolution of the fields is given by the equations 
\beq
\ddot{\sx} + 3 H \dot{\sx} 
= &~ 
- \frac{\partial V}{\partial\sx},
\label{threea} \\
\ddot{\phi}_{1,2} + \left(3 H + \Gamma\right)\dot{\phi}_{1,2} 
= &~ 
- \frac{\partial V}{\partial\phi_{1,2}},
\label{threeone} \\
\ddot{\psi}_{1,2} + 3 H \dot{\psi}_{1,2} = &~ 
- \frac{\partial V}{\partial\psi_{1,2}} 
- \kx^2 \sqrt{\frac{5}{6 \pi^2}}  \sqrt{\rho_R}~ \psi_{1,2},
\label{threetwo} \\
\dot{\rho}_R + 4 H \rho_R = &~
\Gamma \left( \langle \dot{\phi}_1^2 \rangle
+ \langle \dot{\phi}_2^2 \rangle \right),
\label{threethree} \\
H^2 = \left( \frac{\dot{R}}{R} \right)^2 = &~\frac{1}{3 \mpl^2} \left(
\sum_i \frac{1}{2} \dot{\varphi}_i^2 
+ V + \rho_R \right). 
\label{threefour} \eeq
The quantities  
$\langle \dot{\phi}_{1,2}^2 \rangle$ correspond to the average energy 
of oscillations of the $\phi_{1,2}$ fields over one cycle \cite{ct}. 
The gas of particles with energy density $\rho_R$ 
consists of scalar and fermionic $\Psi$ particles 
with mass $2 m$. 
The second term in the right-hand side of eq. (\ref{threetwo}) 
results from the interaction of the $\psi_{1,2}$ fields with the
gas of scalar $\Psi$ particles (see below). 

We have integrated numerically the above equations for
the potential of eq. (\ref{three}) 
with 
$\kx=0.5$, $\lx=0.3$, $\mu/\mpl=10^{-3}$, $m/\mpl=10^{-2}$.
The initial values of the fields have been chosen
$\sx/\mpl=3$,
$\pha/\mpl=0.6$,
$\phb/\mpl=0.5$,
$\psa/\mpl=0.1$,
$\psb/\mpl=0.2$,
They are near $\mpl$, with 
$\pha$, $\phb$, $\psa$, $\psb$ 
smaller than $\sx$, so that the evolution starts
near the flat direction. The parameters of the $\sx$, $\phi_{1,2}$
system have been chosen equal to those of the previous section (see fig. 1). 
The initial time derivatives of the fields
have been taken equal to zero.
The initial energy density is $\rho_0/\mpl^4 \simeq 0.512$ and the 
initial value of the Hubble parameter $H_0/\mpl \simeq 0.413$.

In fig.2 we present the evolution of the field $\pha$. 
We observe that it settles quickly in a pattern of regular oscillations
around zero. The frequency of oscillations is set by the mass
in the first of eqs. (\ref{four}). 
The amplitude of the oscillations decays quickly
because of the large initial values of the Hubble parameter and the
decay rate of eqs. (\ref{six}).
After a time $t\mpl \simeq 5 \times 10^2$, the evolution of $\pha$ is
influenced by its interaction with the oscillating $\psi_{1,2}$ fields
through the last term in the third line of eq. (\ref{three}).
As a result, $\pha$ develops an oscillatory mode with 
characteristic period $\simeq \pi/m$.
In fig. 2 we also present the evolution of the $\sx$ field. 
We observe that it rolls slowly towards the origin because of the 
effective mass generated by the first term in the third line of 
eq. (\ref{three}). As the amplitude of the $\phi_{1,2}$ oscillations
drops, the time derivative 
of $\sx$ is reduced because of the ``friction'' term
$3 H \dot{\sx}$ in its equation of motion.  

The oscillating $\phi_{1,2}$ fields produce an equal number of
scalar and fermionic $\Psi$ particles. 
The scalar 
$\Psi$ particles have a large self-interaction rate and
their gas is expected to reach thermal equilibrium quickly. However, the
interactions of the fermionic $\Psi$ particles involve additional 
heavy $\Phi$ particles and their superpartners,
and thermalization is not possible for them. 
The temperature $T$ of the scalar $\Psi$ gas
is given by 
\be
\rho_{Rs}= \frac{\rho_R}{2} = g_* \frac{\pi^2}{30} T^4,
\label{rhot1} \ee
with an effective number of degrees of freedom $g_*=2.$
The interaction of the $\psi_{1,2}$ fields with the thermal
bath of the scalar $\Psi$ particles is established through the term
$\kx^2 \left( \psa^2 + \psb^2 \right)^2 / 4$ in the potential. 
It results in an effective thermal mass for these fields
\be
m^2_{\psi}(T) = \frac{\kx^2}{3} T^2 = 
\kx^2 \sqrt{\frac{5}{6 \pi^2}}  \sqrt{\rho_R}.
\label{thmass} \ee
We have included this contribution in the equations of motion
(\ref{threetwo}), as it can be much larger than the 
zero-temperature mass $2m$. 
The interactions of the $\psi_{1,2}$ fields with the
fermionic $\Psi$ particles are negligle, as they 
involve 
heavy $\Phi$ particles and their superpartners which are not
expected to be abundant.

In fig. 3 we present the evolution of all 
the fields during the time between 
$t\mpl = 10^3$ and $t\mpl = 2 \times 10^3$. 
After $t\mpl \simeq  10^3$, the fields 
$\phi_{1,2}$, $\psi_{1,2}$ follow a pattern of oscillations with 
characteristic frequency set by the mass of the 
$\psi_{1,2}$ fields $M_{\psa}=M_{\psb}=2m=0.02~\mpl$.
The small 
oscillations of $\phi_{1,2}$, with high characteristic 
frequency set by their mass in eqs. (\ref{four}), quickly die out. 
The amplitude of the low-frequency oscillations 
is reduced because of the ``friction'' term $\sim 3H$ in the equations
of motion. The Hubble parameter is dominated by the 
energy density $\rho_R$ of the gas of $\Psi$ particles. 
At $t_1/\mpl = 1.5 \times 10^3$ this energy density is 
$\left[ \rho_R\right]_1/\mpl^4 \simeq 3.28 \times 10^{-7}$, while the
(potential and kinetic) energy density of the fields is
$\left[ \rho_f \right]_1/\mpl^4 \simeq 5.86 \times 10^{-9}$.
The Hubble parameter is $H_1/\mpl \simeq 3.34 \times 10^{-4}$.
The field $\sx$ has a value $\sx_1/\mpl \simeq 0.978$ and
a time derivative $\dot{\sx}_1/\mpl^2 \simeq -7.26 \times 10^{-5}$.

We have found numerically that, in the time interval between 
$t\mpl \simeq 10^3$ and $t\mpl \simeq 1.5 \times 10^3$, 
the (kinetic and potential) energy density of the fields
$\rho_f$ falls proportionally to a power $\gta 4$ of the scale factor $R$. 
The reason is that the dynamics of 
the fields during this time is strongly affected by the
quartic terms in the potential and there is still significant decay of
the $\phi_{1,2}$ fields into $\Psi$ particles. 
At times $t\mpl \gta 1.5 \times 10^3$, 
$\rho_f$ falls proportionally to approximately
the third power of $R$.
This is due to the fact that the amplitude of the field
oscillations becomes sufficiently small for the quartic terms to
be irrelevant, while the particle production practically stops. 
The radiation energy density falls as
$R^{-4}$ until a time 
$t\mpl \sim 2 \times 10^3$. At this point 
the typical energy of a scalar $\Psi$ particle $E \sim T$
is expected to be comparable to its mass. Similar conlusions can be
reached for the typical energy of a  
fermionic $\Psi$ particle. Based on the above considerations, we assume that, 
after $t\mpl \gta 1.5 \times 10^3$, the Universe becomes
matter dominated and the energy density of the 
particle gas falls as  $R^{-3}$, similarly to the
field energy density. As a result, the ratio of 
the field energy density $\rho_f$ to the 
radiation energy density $\rho_R$ stays constant 
$e_f \sim \left[ \rho_f \right]_1/\left[ \rho_R \right]_1
\simeq 1.79 \times 10^{-2}$.

The evolution of the $\sx$ field after $t_1/\mpl = 1.5 \times 10^3$ 
is very simple. The right-hand side of eq. (\ref{threea}) is 
negligible compared to the ``friction'' term proportional to
$3H$ in the left-hand side. 
The Hubble parameter as a function of time is given by 
$H = 2/\left[3(1+w)t\right]$.
The parameter $w$ determines the relation between energy density and
the mean value of the pressure ($p=w\rho)$. 
For a system of massless oscillating fields, or a 
radiation-dominated Universe,
$w = 1/3$. 
For a system of massive oscillating fields, or a 
matter-dominated Universe (such as the case we are considering),
$w = 0$.
The solution of eq. (\ref{threea})
is given by 
\be
\sx(t) = \sx_1 + \dot{\sx}_1 \frac{1+w}{1-w}~
t_1^{\frac{2}{1+w}} \left(
t_1^{\frac{-1+w}{1+w}}
- t^{\frac{-1+w}{1+w}}
\right).
\label{laone} \ee
Asymptotically, the $\sx$ field is expected to settle down at a 
value $\sx_{f}/\mpl = 0.869$.
We have verified numerically the evolution of $\sx$ according to 
eq. (\ref{laone}) and that the Universe is
always dominated by the energy density of 
the gas of $\Psi$ particles.

Inflation is expected to start at a time 
$t \mpl \sim 10^6$, when $\rho_R \sim \mu^4$. 
The size of the amplitudes of the 
$\phi_{1,2}$ field oscillations at this point 
is crucial for the onset of inflation. They must satisfy 
eq. (\ref{twosixteen}) for $\sx$  not to be disturbed 
from its constant value on the inflationary valley of the potential.   
The total energy density of the fields remains smaller 
than the radiation energy density 
by a factor 
$e_f  \simeq 1.79 \times 10^{-2}$. 
Moreover, the largest fraction of this energy density is
stored in oscillations of the $\psi_{1,2}$ fields, whose
amplitude is larger by a factor $\sim 10^2$ than that of
the $\phi_{1,2}$ fields. It is easy then to convince oneself
that the constraint of eq. (\ref{twosixteen}) is comfortably
satisfied at the onset of inflation. 
Moreover, as we discuss in the following section, 
the low-frequency oscillations of the $\phi_{1,2}$ fields
do not generate an effective mass term for $\sx$. Only the
high-frequency ones contribute to this effective mass, and they
are much more suppressed at a time $t \mpl \sim 10^6$.
We conclude 
that inflation sets in as soon as the energy density
falls below $\mu^4$, without the requirement of a
fine-tuned configuration at the end of 
the Planck era.   

\section{Discussion} 

It is possible to obtain an approximate analytical 
solution for the field evolution during the later stages of 
particle production by the oscillating $\phi_{1,2}$ fields.
At the end of the Planck era the 
energy density is $\sim \mpl^4$ 
and the ``friction'' term in the equations of motion is
dominated by the contribution proportional to $3H$.
However, after a few oscillations of $\phi_{1,2}$
the Hubble parameter becomes smaller than $\Gamma$.
The amplitude of the oscillations falls exponentially with
time as the $\sx$ field varies very slowly. 
After a time $t \mpl \sim 10^2$ 
the amplitude of $\phi_{1,2}$ has dropped sufficiently for the 
evolution equations to be approximated as 
\be 
\ddot{\phi}_{1,2} + \Gamma \dot{\phi}_{1,2} + M^2_\phi \phi_{1,2} 
= f_{1,2}(t),  
\label{x} \ee 
with $M^2_{\phi_{1,2}}\simeq M^2_\phi = 2\lx^2\sx^2$. 
The source terms $f_{1,2}$ are determined by the last term in 
the third line of eq. (\ref{three}):
$f_1 = -\kappa \lx \sx(\psi_1^2 - \psi_2^2)$, 
$f_2 = -2\kappa \lx \sx\psi_1 \psi_2$. The $\sx$ field is almost 
constant, while the $\psi_{1,2}$ fields
oscillate with a characteristic period $\sim
\pi/m \simeq 300~\mpl^{-1}$, 
much larger than the period $\sim \pi/\mpl$ 
of the $\phi_{1,2}$ oscillations.  

An approximate solution of eqs. (\ref{x}) is given by 
\be
\phi_{1,2}(t) = \bar{ \phi}_{1,2}(t) + \delta \phi_{1,2}(t) =
\frac{f_{1,2}(t)}{M_{\phi}^2}
+\phi_0 e^{-\frac{\Gamma}{2} t} \cos \left( M_{\phi} t + a \right). 
\label{phi} \ee 
The term $\delta \phi_{1,2}$ of this solution describes the decaying 
high-frequency oscillations, while the term $\bar{ \phi}_{1,2}$
determines the variation of $\phi_{1,2}$ induced by 
the slower evolution of $\psi_{1,2}$.
It can easily be checked that, for $\phi_{1,2} = \bar{ \phi}_{1,2}$,
the potential becomes independent of $\sx$.
This implies that the effective slope of $\sx$ during the later
stages of the evolution is generated only by  
the high-frequency oscillations 
$\delta \phi_{1,2}(t)$. It is their amplitudes that must 
satisfy the bound of eq. (\ref{twosixteen}). Due to the exponential
decay of these oscillations until the onset of inflation, 
their amplitudes are smaller than the upper limit allowed
by eq. (\ref{twosixteen}) by several orders of magnitude.

A few more remarks are in order with respect to
the $\sx$-field evolution before the onset of inflation.
Its behaviour is not significantly 
affected by the presence of the gas of $\Psi$ particles. 
At early times its evolution 
is governed by the interaction with the oscillating
$\phi_{1,2}$ fields. The effective interaction 
with the thermal gas of scalar $\Psi$ particles
can be described by the last 
term in the third line of eq. (\ref{three}) 
if the products involving $\psi_{1,2}$ fields are
replaced by their thermal averages. 
The resulting contribution is zero. 
The interactions with the gas of 
fermionic $\Psi$ particles involve massive $\Phi$ 
particles, which are not expected to be abundant. 
As a result, the $\sx$ field is insensitive to the 
presence of the fermionic $\Psi$ particles. 

The supergravity corrections are not expected to modify the above picture
significantly.
The expansion parameters for these corrections are $\sx^2/2\mpl^2$,
$\phi_{1,2}^2/2\mpl^2$, $\psi_{1,2}^2/2\mpl^2$. For our model
high powers of 
$\sx^2/2\mpl^2$ are significant, but these are multiplied by 
powers of $\mu^2/\mpl^2$ near the flat direction  
$\phi_{1,2}=\psi_{1,2}=0$. 
The most important supergravity correction for our discussion
arises when the term $\kx^2 \left( \psi_1^2+\psi_2^2 \right)^2/4$ 
in the potential is multiplied
by a correction $\sx^2/(2\mpl^2)$ 
originating in the K\"ahler potential.
A contribution to the 
effective mass of $\sx$ is obtained if the 
thermal average of the $\left( \psi_1^2+\psi_2^2 \right)^2$ term 
is considered. One obtains \cite{megeorge}
\be
 m^2_{\sx}(T) = f \frac{\rho_R}{\mpl^2},
\label{ouf} \ee
with 
\be 
f \sim
\kx^2 \left( \frac{T^2}{12} \right)^2 \frac{1}{\rho_R}
\lta 10^{-2} .
\label{effst} \ee
The evolution of $\sx$ as a function of the
energy density is determined by the Euler equation \cite{megeorge}
\be
\rho_R^2~ \frac{d^2 \sx}{d\rho_R^2} 
+ \left( \frac{3}{2} - \frac{1}{1+w} \right) \rho_R~\frac{d \sx}{d\rho_R}
+ \frac{f}{3(1+w)^2} ~\sx =0,
\label{llaone} \ee
with the approximate solution 
\be
\sx = \sx_i \left( \frac{\rho_R}{\rho_{Ri}} 
\right)^{\frac{2f}{3\left( 1-w^2 \right)}}.
\label{appr} \ee
Here $\rho_{Ri}/\mpl^4 \sim 10^{-5}$,
$\sx_i \sim 1$ correspond to the time when the term 
of eq. (\ref{ouf}) starts to dominate over the 
mass term $\sim \lx^2 \left( \pha^2 + \phb^2 \right)$ from 
the potential of eq. 
(\ref{three}).
The 
decrease of $\sx$ generated by this supergravity-induced term
until the onset of inflation 
is expected to be  $\Delta \sx/\mpl \sim 0.2$ and 
should not affect the qualitative features of our study.

We conclude that the decay of the fields orthogonal to the
inflaton into $\Psi$ particles
leads to the onset of hybrid inflation without
the need of an extremely homogeneous initial field configuration.
The gas of $\Psi$ particles produced before the inflationary 
era is diluted by the subsequent expansion and their number
density is negligible today. 

It should be pointed out, however, that a certain degree of
homogeneity is still necessary. The field evolution must 
start closer to the flat direction of the potential 
than to its minima. According to the discussion at the end 
of the first section, this requirement must be satisfied
over distances 
far beyond the initial Hubble length. 
This means that some initial homogeneity must exist and extend 
far beyond the typical size of regions that can be considered
causally connected. 
This problem can be resolved if inflation starts at
an energy scale near $\mpl$, as for example in the models
of two-stage inflation of refs. \cite{mecostas,megeorge}.

\vspace{0.5cm}
\noindent
{\bf Acknowledgements}: 
We would like to thank 
A. Dolgov, G. Dvali, D. Kirilova, and A. Romanino 
for useful discussions.
We would also like to thank the organizers of the Extended Workshop
on Highlights in Astroparticle Physics at the ICTP, Trieste,  
where this work has been initiated.  
Z.B. thanks the INFN section of Padua and the CERN theory 
division for the hospitality during the last stages of the work. 
N.T. was supported by the E.C.
under TMR contract No. ERBFMRX--CT96--0090.


\newpage

\pagestyle{empty}

\begin{figure}
\psfig{figure=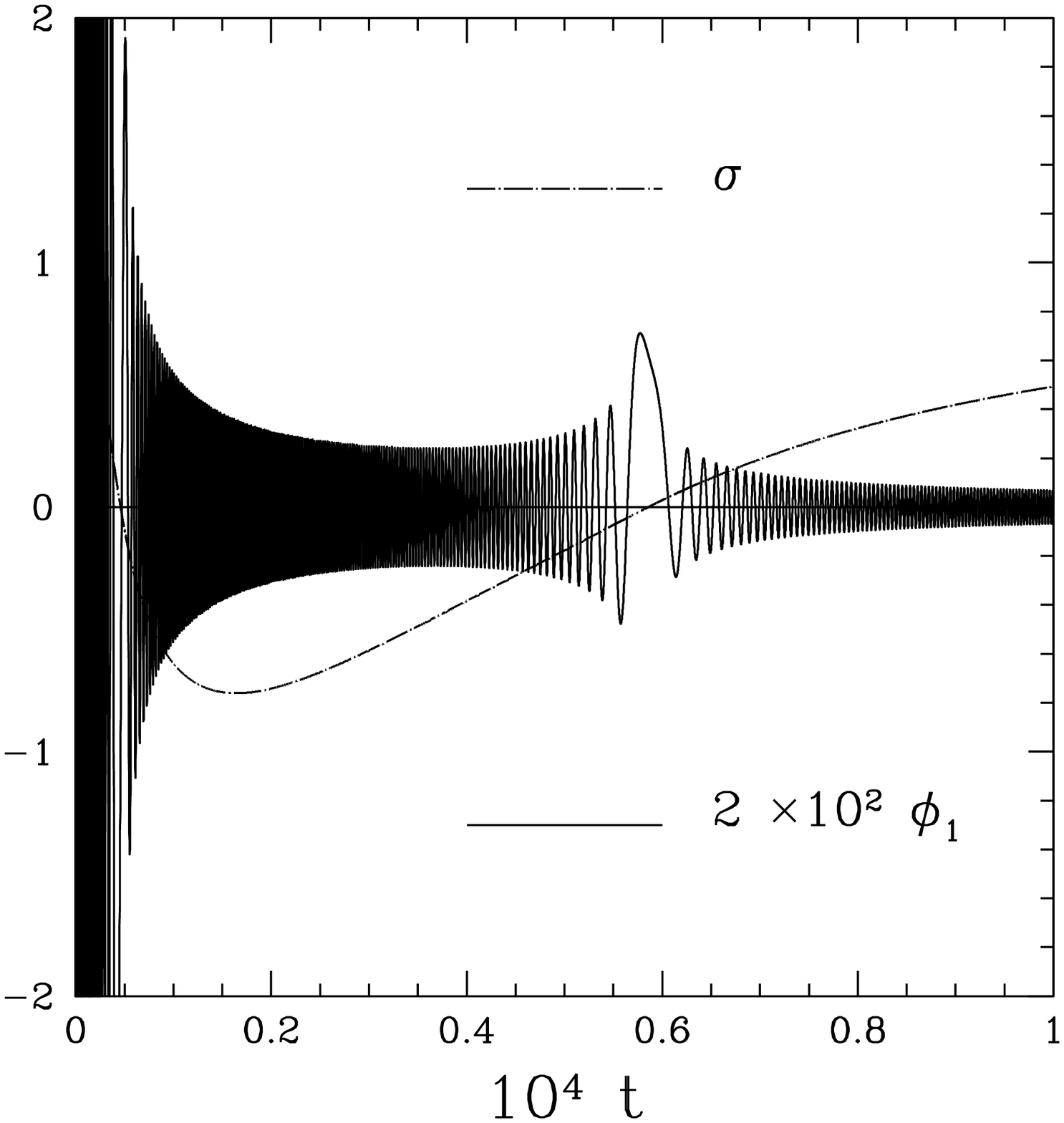,height=16.0cm}
Fig. 1: 
The evolution of 
$\sx$, $\pha$, 
for a theory described by
the potential of eq. (\ref{three1}) 
with 
$\lx=0.3$, $\mu/\mpl=10^{-3}$.
The initial values of the fields have been chosen
$\sx/\mpl=3$,
$\pha/\mpl=0.6$,
$\phb/\mpl=0.5$.
Dimensionful quantities are given in units of $\mpl$.
\end{figure}

\begin{figure}
\psfig{figure=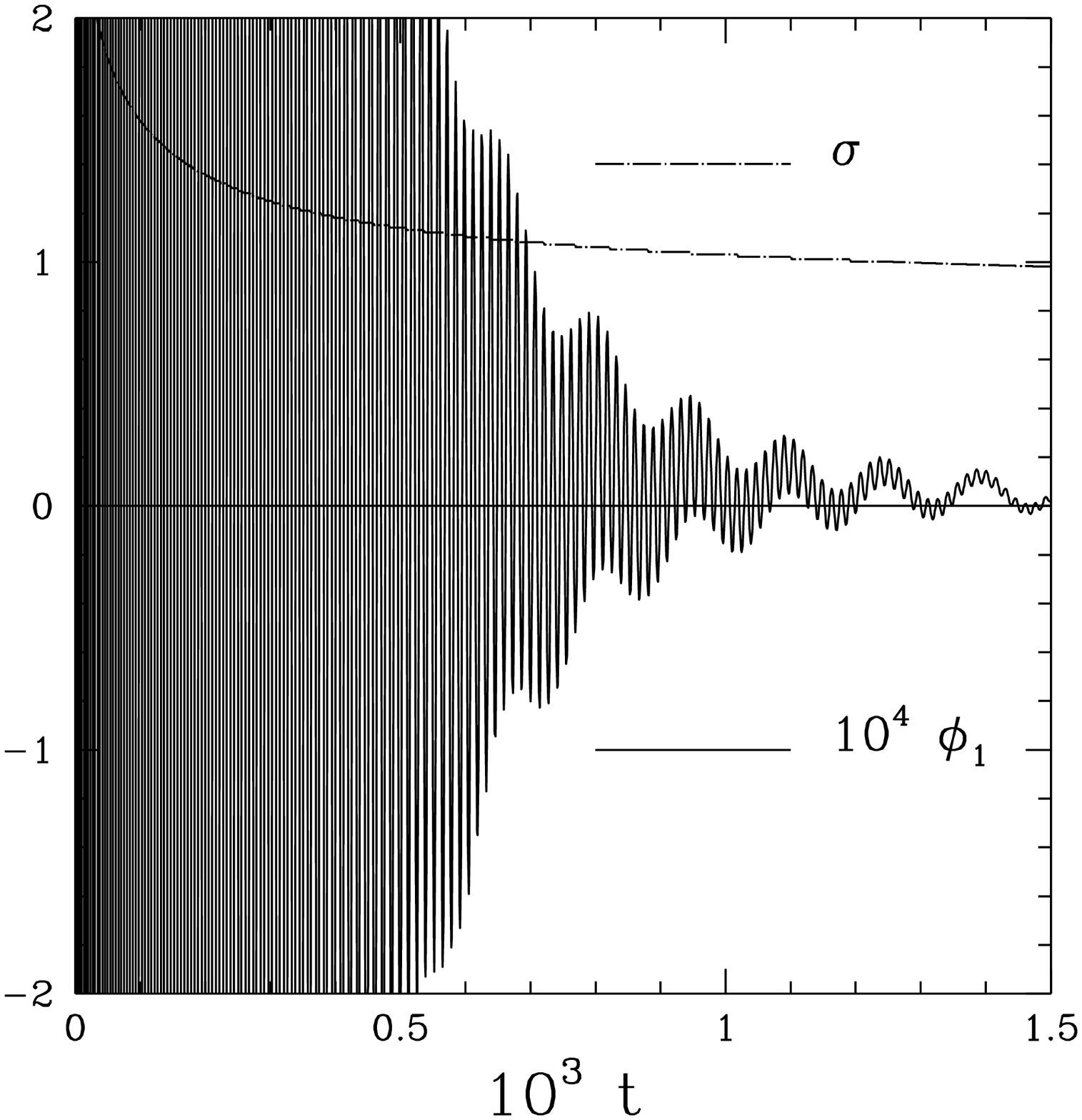,height=16.0cm}
Fig. 2: 
The evolution of 
$\sx$, $\pha$, 
for a theory described by
the potential of eq. (\ref{three}) 
with 
$\kx=0.5$, $\lx=0.3$, $\mu/\mpl=10^{-3}$, $m/\mpl=10^{-2}$.
The initial values of the fields have been chosen
$\sx/\mpl=3$,
$\pha/\mpl=0.6$,
$\phb/\mpl=0.5$,
$\psa/\mpl=0.1$,
$\psb/\mpl=0.2$.
Dimensionful quantities are given in units of $\mpl$.
\end{figure}

\begin{figure}
\psfig{figure=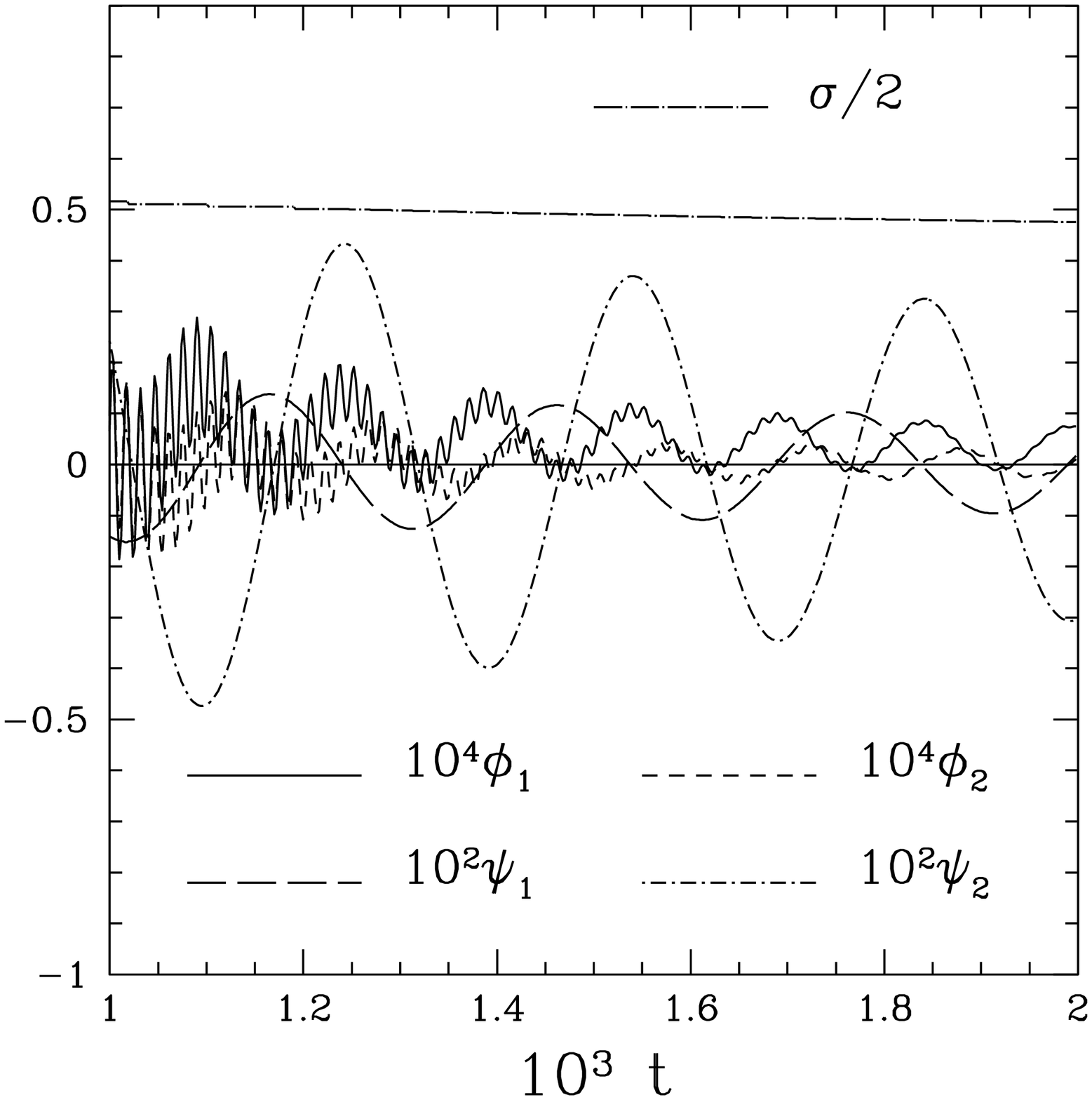,height=16.0cm}
Fig. 3: 
Same as in fig. 2 for 
$\sx$, $\pha$, $\phb$, $\psa$, $\psb$.
\end{figure}

\end{document}